\documentclass[10pt,twoside]{article}
\usepackage{Latex-arXiv}
\almvol{00}
\almttone{Frontiers of Science Awards}
\almtttwo{for Math/TCIS/Phys}
\firstpage{1}
\usepackage[english]{babel}
\usepackage[utf8]{inputenc}

\usepackage{epsfig}
\usepackage{axodraw2}

\def\bubbleonenum#1#2{
\raisebox{-7pt}
{
\begin{axopicture}{(30,20)(-13,-10)}
\SetScale{1}\SetColor{Blue}%
\Line(-15,0)(-10,0)
\Line(10,0)(15,0) 
\CArc(0,0)(10,0,360)
\Vertex(-10,0){1.5}
\Vertex(10,0){1.5}
\SetScale{1}\SetColor{Black}%
\Text(1.5,6.5){\tiny $#1$ \tiny}
\Text(1.5,-6.5){\tiny $#2$ \tiny}
\end{axopicture}
}
}

\def\bubtwodotnum#1#2#3{
\raisebox{-7pt}
{
\begin{axopicture}{(30,20)(-13,-10)}
\SetScale{1}\SetColor{Blue}%
\Line(-15,0)(-10,0)
\Line(10,0)(15,0) 
\CArc(0,0)(10,0,360)
\Line(-10,0)(10,0)
\Vertex(-10,0){1.5}
\Vertex(10,0){1.5}
\Vertex(0,10){1.5}
\SetScale{1}\SetColor{Black}%
\Text(1,14){\tiny $#1$ \tiny}
\Text(1,3){\tiny $#2$ \tiny}
\Text(1,-7){\tiny $#3$ \tiny}
\end{axopicture}
}
}

\def\IRonenum#1{
\raisebox{-7pt}
{
\begin{axopicture}{(10,20)(-4,-10)}
\SetScale{1}\SetColor{RedViolet}%
\Line(0,-10)(0,10)
\CCirc(0,-10){1.5}{RedViolet}{White}
\Vertex(0,0){1.5}
\CCirc(0,10){1.5}{RedViolet}{White}
\SetScale{1}\SetColor{Black}%
\Text(6,0){\tiny $#1$ \tiny}
\end{axopicture}
}
}

\def\vaconenum#1{
\raisebox{-7pt}
{
\begin{axopicture}{(30,20)(-13,-10)}
\SetScale{1}\SetColor{Blue}%
\CArc(0,0)(10,0,360)
\Vertex(-10,0){1.5}
\Vertex(10,0){1.5}
\SetScale{1}\SetColor{Black}%
\Text(1.5,-6.5){\tiny $#1$ \tiny}
\end{axopicture}
}
}

\def\tadonenum#1{
\raisebox{-7pt}
{
\begin{axopicture}{(30,20)(-13,-10)}
\SetScale{1}\SetColor{Blue}%
\CArc(0,5)(5,0,360)
\Line(-10,0)(10,0)
\Vertex(0,0){1.5}
\Vertex(0,10){1.5}
\SetScale{1}\SetColor{Black}%
\Text(1.5,5){\tiny $#1$ \tiny}
\end{axopicture}
}
}

\usepackage{amsmath}
\usepackage{amssymb}
\usepackage{cite}

\usepackage{slashed}
\usepackage{hyperref}
\usepackage{tikz}

\numberwithin{equation}{section}

\newcommand{\beq}{\begin{equation}}
\newcommand{\eeq}{\end{equation}}
\newcommand{\bea}{\begin{eqnarray}}
\newcommand{\eea}{\end{eqnarray}}
\newcommand{\hspn}{{\hspace{-4mm}}}

\newcommand{\nn}{\nonumber}
\newcommand{\MSb}{$\overline{\mbox{MS}}$}
\newcommand{\ra}{\rightarrow}

\newcommand{\gsim}{\raisebox{-0.07cm}{$\:\stackrel{>}{{\scriptstyle
 \sim}}\: $} }
\newcommand{\lsim}{\raisebox{-0.07cm}{$\:\stackrel{<}{{\scriptstyle
 \sim}}\: $} }

\newcommand{\als}{\alpha_{\rm s}}
\newcommand{\ars}{a_{\rm s}}

\newcommand{\muf}{\mu_{\:\!\!f}^{}}
\newcommand{\mufs}{\mu_{\:\!\!f}^{\,2}}

\newcommand{\ep}{\varepsilon}

\begin{document}

\def\z#1{{\zeta_{#1}}}
\def\zss{\zeta_2^{\,2}}

\def\nc{{n_c}}
\def\ncs{{n_{c}^{\,2}}}
\def\nct{{n_{c}^{\,3}}}

\def\ca{{C^{}_{\!A}}}
\def\cas{{C^{\,2}_{\!A}}}
\def\cat{{C^{\,3}_{\!A}}}
\def\caf{{C^{\,4}_{\!A}}}
\def\cai{{C^{\,5}_{\!A}}}

\def\cf{{C^{}_{\!F}}}
\def\cfs{{C^{\, 2}_{\!F}}}
\def\cft{{C^{\, 3}_{\!F}}}
\def\cff{{C^{\, 4}_{\!F}}}

\def\nf{{n^{}_{\! f}}}
\def\nfz{{n^{\,0}_{\! f}}}
\def\nfo{{n^{\,1}_{\! f}}}
\def\nfs{{n^{\,2}_{\! f}}}
\def\nft{{n^{\,3}_{\! f}}}
\def\nff{{n^{\,4}_{\! f}}}

\def\tf{{T^{}_{\!F}}}
\def\tfs{{T^{\,2}_{\!F}}}
\def\tft{{T^{\,3}_{\!F}}}
\def\tff{{T^{\,4}_{\!F}}}

\def\dfAAna{{\frac{d_A^{\,abcd}d_A^{\,abcd}}{N_A }}}
\def\dfFAna{{\frac{d_F^{\,abcd}d_A^{\,abcd}}{N_A }}}
\def\dfFFna{{\frac{d_F^{\,abcd}d_F^{\,abcd}}{N_A }}}

\def\nl{{n^{}_{\! f}}}
\def\nlz{{n^{\:\!0}_{\! f}}}
\def\nlo{{n^{\:\!1}_{\! f}}}
\def\nls{{n^{\:\!2}_{\! f}}}
\def\nlt{{n^{\:\!3}_{\! f}}}
\def\nlf{{n^{\:\!4}_{\! f}}}

\def\as(#1){{\alpha_{\rm s}^{\:#1}}}
\def\ar(#1){{a_{\rm s}^{\:#1}}}

\def\MZ{{M_{\rm Z}}}
\def\MZs{{M_{\rm Z}^{\:\!2}}}

\def\MHt{{M_{\rm H}^{\:\!3}}}
\def\MHs{{M_{\rm H}^{\:\!2}}}
\def\MH{{M_{\rm H}}}

\def\frak#1#2{\mbox{\large{$\frac{#1}{#2}$}}}
\def\frct#1#2{\mbox{\footnotesize{$\displaystyle\frac{#1}{#2}$}}}

\def\muRs{{\mu_R^{\,2}}}
\def\L{\mathcal{L}}
\def\eps{\epsilon}
\def\dots{..}
\def\r{\right}
\def\l{\left}

\markboth{\hfill
{\rm F. Herzog, B. Ruijl, T. Ueda, J. Vermaseren and A. Vogt} \hfill}
{\hfill {\rm Five-loop beta function for gauge theories \hfill}}

\title{Five-loop beta function for gauge theories:$\!\!\!$\\[-0.5mm] 
computations, results and consequences}

\author{$\,$\\ Franz Herzog, Ben Ruijl, Takahiro Ueda%
\footnote{Speaker at the 2025 International Conference of Basic Science, 
presenting the article \cite{Herzog:2017ohr} which received a 2025 
Frontier of Science Award in Theoretical Physics.}$\!$,\\[1mm] 
Jos Vermaseren and Andreas Vogt}

\begin{abstract}
At the end of 2016, we computed the five-loop (N$^4$LO) contributions 
to the beta function in perturbative Quantum Chromodynamics (QCD), 
its generalization to non-Abelian gauge theories with a simple compact 
Lie group, and for Quantum Electrodynamics (QED).
Here we recall main tools used in and specifically developed 
for this computation and its main analytic and numerical results.
The development work carried out for this project facilitated further 
even more involved analytic five-loop computations.
We briefly summarize also their numerical QCD results for Higgs-boson 
decay to hadrons in the heavy-top limit and for two N$^4$LO splitting 
functions for the evolution of quark distributions of hadrons. 
The latter lead to a first realistic estimate of the five-loop 
contribution to another important quantity in perturbative QCD, 
the quark cusp anomalous dimension.
\end{abstract}

\maketitle


\section{Introduction}

The beta function, which governs the scale dependence of the renormalized
coupling constant, is arguably the most important fundamental property
of interacting quantum field theories.
For QCD and its generalization to other non-Abelian gauge theories, this
function has been known for almost three decades to four-loop
(next-to-next-to-next-to-leading order, N$^3$LO) accuracy
\cite{Gross:1973id,Politzer:1973fx,Caswell:1974gg,Jones:1974mm,%
Tarasov:1980au,Larin:1993tp,vanRitbergen:1997va,Czakon:2004bu}.
By now, this order has become the accuracy frontier for the analyses of 
benchmark quantities, such as the cross section for Higgs-boson production, 
at the LHC.

\vspace{1mm}
A little less than 10 years ago, several groups undertook to extend the
above results to five loops. First the results for the gauge group SU$(3$), 
i.e., QCD were obtained in ref.~\cite{Baikov:2016tgj}. 
Their result did not meet all theoretical expectations \cite{Shrock-pc} in 
the context of ref.~\cite{Ryttov:2016asb}. 
We verified the result of ref.~\cite{Baikov:2016tgj} and provided its
extension to a general simple compact Lie group in ref.~\cite{Herzog:2017ohr}.
Two more determinations, performed soon thereafter by different means,
confirmed our results {\cite{Luthe:2017ttg,Chetyrkin:2017bjc}}. 

\vspace{1mm}
Unlike the other computations, ours used the background field method
\cite{Abbott80,AbbottGS83} and infrared rearrangement via a newly
developed diagram-by-diagram implementation \cite{Herzog:2017bjx}
of the R$^*$ operation \cite{RSTAR1982,RSTAR1984,RSTAR1985,RSTAR1991} 
that allowed to compute the pole terms 
of the required five-loop diagrams in terms of four-loop propagator-type
integrals. These were evaluated with the then new {\sc Forcer} program
\cite{Ruijl:2017cxj} in {\sc Form} 
\cite{Vermaseren:2000nd,Kuipers:2012rf,Ruijl:2017dtg}. 
Thanks to the tools developed for the determination of the 5-loop
beta functions, we were later able to compute other five-loop quantities
requiring even harder calculations, mostly due to the presence of
high-rank tensor integrals \cite{Herzog:2017dtz,Herzog:2018kwj}.

\vspace{1mm}
The remainder of this article is organized as follows.
In section 2 we discuss the above-mentioned techniques employed for 
ref.~\cite{Herzog:2017ohr}. 
We then turn to the analytic and numerical results for the N$^4$LO 
beta function of QCD and its generalization in section 3. 
A shorter discussion of our other five-loop results is provided in 
section~4, 
before we close with a brief summary in section 5.

\section{Concepts, codes and computations}

\subsection{The background field method}
 
A convenient and efficient method to extract the Yang-Mills beta function 
is the background field gauge, which we review in the following.
The Lagrangian of Yang-Mills theory coupled to fermions in a non-trivial 
(often the fundamental) representation of the gauge group, the theory for 
which we will present the five-loop beta-function in the next section,
can be decomposed as
\beq
  \L_{\text{YM+FER}} \;=\;
  \L_{\text{CYM}}+\L_{\text{GF}}
 +\L_{\text{FPG}}+\L_{\text{FER}}\,.
\eeq
Here the classical Yang-Mills Lagrangian (CYM), a gauge-fixing term (GF),
the Faddeev-Popov ghost term (FPG) and the fermion term (FER) are given by
\bea
\L_{\text{\sc CYM}} &\,=\,& -\frac{1}{4}\, F_{\mu\nu}^a(A)F^{\mu\nu}_a(A)
\, , \;\; 
\L_{\text{\sc GF}}  \;=\; -\frac{1}{2\xi}\, (G^a)^2
\,, \nn \\[1mm]
\L_{\mathrm{\sc FPG}} &\,=\,& -\eta^\dagger_a \,\partial^{\:\!\mu} D^{ab}_\mu(A)
  \,\eta_b^{}
\, , \quad
 \L_{\mathrm{\sc FER}} \;=\; \sum_{i,j,f}^{}\bar\psi_{if}^{}(i\slashed{D}_{ij}(A)
  -m_{\!f}^{}\delta_{ij})\, \psi_{jf}^{}\, . 
\quad
\eea
In the fermionic term the sum goes over colours $i,j$, and $\nf$ flavours $f$,
and we employ the standard Feynman-slash notation. 
The field strength is defined
by
\beq
\quad F_{\mu\nu}^a(A) \;=\; \partial_{\:\!\mu} A_\nu^a
  -\partial_{\:\!\nu} A_\mu^a + g f^{abc} A^b_{\mu}A^c_{\nu}
\, ,
\eeq
and the covariant derivatives are given by
\beq
D^{ab}_\mu(A) \,=\, \delta^{\:\!ab}\partial_\mu -g f^{abc} A^c_\mu
\, ,  \quad
D_{ij}^\mu(A) \,=\, \delta_{ij}\partial^{\:\!\mu} -ig\, T^{a}_{ij} A_a^\mu
\, .
\eeq
The $T^a$ are the generators and the $f^{abc}$ the structure constants of 
a compact gauge group, satisfying the Lie algebra $[T^a,T^b]=f^{abc} T^c$.
The gauge-fixing term is usually chosen as 
$G^a=\,\partial^{\:\!\mu} A_\mu^a$.

The background-field Lagrangian is derived by expressing the gauge field as
\beq
\label{bfg}
A_\mu^a(x) \;=\; B_\mu^{\,a}(x)+\hat A_\mu^a(x)\,,
\eeq
where $B_\mu^{\,a}(x)$ is the \emph{classical} background field while $\hat
A_\mu^a(x)$ contains the \emph{quantum} degrees of freedom of the gauge
field $A_\mu^a(x)$. The Lagrangian is then
\beq
\label{eq:BYM}
\L_{\mathrm{BYM+FER}} \;=\; \L_{\mathrm{BCYM}}+\L_{\mathrm{BGF}}
  +\L_{\mathrm{BFPG}}+\L_{\mathrm{BFER}}\,.
\eeq
$\L_{\mathrm{BCYM}}$ and $\L_{\mathrm{BFER}}$ are derived similarly by
substituting eq.~(\ref{bfg}) into the corresponding terms in the Yang-Mills
Lagrangian. However, a clever choice exists \cite{Abbott80,AbbottGS83} for the
ghost and gauge-fixing terms, which allows this Lagrangian to maintain explicit
gauge invariance for the background field $B_\mu^a(x)$, while fixing only the
gauge freedom of the quantum field  $\hat A_\mu^a(x)$:
\beq
G^a \;=\; D_\mu^{ab}(B) \hat A^\mu_{b}\, .
\eeq
The ghost term (which follows from BRST symmetry) is given by
\beq
\L_{\mathrm{BFPG}} \;=\;  -\eta^\dagger_a \, D^{ab;\mu}(B) \,
  D^{bc}_\mu(B+\hat A) \,\eta_c\, .
\eeq
The Lagrangian $\L_{\mathrm{BYM+FER}}$ gives rise to additional
interactions which are different from the normal QCD interactions of the
quantum field $\hat A_\mu^a(x)$ since they also contain interactions of 
$B_\mu^{\,a}(x)$ with all other fields.

\begin{figure}
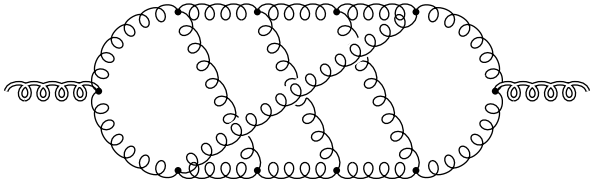

\vspace*{-1mm}
\begin{center}
\begin{axopicture}(230,80)(0,0)
\Gluon(130,70)(160,10){3}{8}
\Gluon(100,70)(130,10){3}{8}
\Gluon(70,70)(100,10){3}{8}
\Line[color=white,width=8](160,70)(70,10)
\DoubleGluon(5,40)(40,40){3}{4}{1.3}
\GluonArc(70,40)(30,90,180){3}{6}
\GluonArc(70,40)(30,180,270){3}{6}
\Gluon(160,70)(70,10){3}{15}
\Gluon(100,70)(70,70){-3}{4}
\Gluon(130,70)(100,70){-3}{4}
\Gluon(160,70)(130,70){-3}{4}
\Gluon(130,10)(160,10){-3}{4}
\Gluon(100,10)(130,10){-3}{4}
\Gluon(70,10)(100,10){-3}{4}
\GluonArc(160,40)(30,270,360){3}{6}
\GluonArc(160,40)(30,0,90){3}{6}
\DoubleGluon(190,40)(225,40){3}{4}{1.3}
\Vertex(40,40){1.3}
\Vertex(190,40){1.3}
\Vertex(70,70){1.3}
\Vertex(100,70){1.3}
\Vertex(130,70){1.3}
\Vertex(160,70){1.3}
\Vertex(70,10){1.3}
\Vertex(100,10){1.3}
\Vertex(130,10){1.3}
\Vertex(160,10){1.3}
\end{axopicture}
\vspace{-3mm}
\caption{ \small
One of the more complicated diagrams.  Single lines represent gluons,
and the external double lines represent the background field. The presence of
the 10 purely gluonic vertices creates a large expression after the
substitution of the Feynman rules.}
\label{fig:gluons}
\end{center}
\vspace*{-3mm}
\end{figure}

\vspace*{1mm}
The main advantage of the background field gauge, see e.g., 
refs.~\cite{Abbott80,AbbottGS83}, is that
the coupling renormalization, $g\to Z_g\, g$, which determines the beta
function, is directly related to that of the background field, 
$B\to B Z_B$, via the identity
\beq
Z_g\sqrt{Z_B} \;=\;1\, .
\eeq
In the Landau gauge the only anomalous dimension needed in the
background field formalism is thus the beta function. In the Feynman gauge,
on the other hand, the gauge parameter $\xi$ also requires the renormalization
constant $Z_\xi$ --- which equals the gluon field renormalization constant
--- but only to one loop less. In turn, this allows one to extract the beta
function from the single equation
\beq
\label{eq:PIBfinite}
Z_B(1+\Pi_B(Q^2;Z_\xi\xi,Z_g g)) \;=\; \mathrm{finite}
\eeq
with
\beq
\label{eq:PIBtensor}
\Pi_B^{\mu\nu}(Q;Z_\xi\xi,Z_g g) \;=\;
  (Q^2g^{\mu\nu}-Q^\mu Q^\nu) \: \Pi_B(Q^2;Z_\xi\xi,Z_g g)
\eeq
where $\Pi_B^{\mu\nu}(Q^2;\xi,g)$ is the bare self-energy of the background
field. This self-energy is computed by keeping the fields $B$ external while
the only propagating fields are $\hat A, \eta$ and $\psi$. A~typical diagram
contributing to $\Pi_B(Q^2;\xi,g)$ is given in fig.~\ref{fig:gluons}.

\vspace{1mm}
Obtaining the beta function through the background field
gauge is faster and simpler than the traditional method of
computing the gluon propagator, ghost propagator and ghost-ghost-gluon
vertex due to a lower total number of diagrams and the above reduction to a
scalar renormalization.

\subsection{The $R^*$-operation}
 
As outlined above, the five-loop beta function can be extracted from the poles 
(in the dimensional regulator $\ep$) of the bare background field self-energy 
$\Pi_B(Q)$. 
Despite the fact that the five-loop master integrals have by now been computed 
\cite{Georgoudis:2021onj}, it is still beyond current computational 
capabilities to calculate the required five-loop propagator integrals directly.
The main obstacle is the difficulty of performing the required 
integration-by-parts (IBP) reductions.

\vspace*{1mm}
Fortunately the problem can be simplified with the $R^*$-operation. In 
particular, the $R^*$-operation \cite{RSTAR1982,RSTAR1984,RSTAR1985,RSTAR1991} 
is capable of rendering any propagator integral finite by adding to it a number
of suitable subtraction terms. 
The subtraction terms are built from potentially high-rank tensor subgraphs of 
the complete graph, whose tensor reduction can become very involved and 
presented one of the bottlenecks of the calculation. 

\vspace*{1mm}
To tackle this obstacle a new method to construct projectors with the aid of 
an orbit partition was developed. 
This was summarized briefly in the appendix of ref.~\cite{Herzog:2017ohr}.
In the meantime this approach to tensor reduction has been further refined 
and generalised in ref.~\cite{Goode:2024mci}. 
There also exists a public implementation in FORM \cite{Goode:2024cfy}. 
The projectors have also found application in the more general construction 
of ref.~\cite{Anastasiou:2023koq}.

\begin{figure}
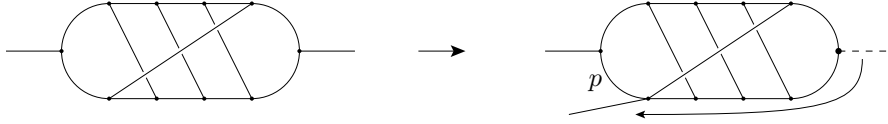

\vspace*{-8mm}
\begin{center}
\begin{axopicture}(390,80)(0,0)
\SetOffset(0,10)
\SetScale{0.6}
\Line(130,70)(160,10)
\Line(100,70)(130,10)
\Line(70,70)(100,10)
\Line[color=white,width=4](160,70)(70,10)
\Line(5,40)(40,40)
\Arc(70,40)(30,90,180)
\Arc(70,40)(30,180,270)
\Line(160,70)(70,10)
\Line(100,70)(70,70)
\Line(130,70)(100,70)
\Line(160,70)(130,70)
\Line(130,10)(160,10)
\Line(100,10)(130,10)
\Line(70,10)(100,10)
\Arc(160,40)(30,270,360)
\Arc(160,40)(30,0,90)
\Line(190,40)(225,40)
\Vertex(40,40){1.3}
\Vertex(190,40){1.3}
\Vertex(70,70){1.3}
\Vertex(100,70){1.3}
\Vertex(130,70){1.3}
\Vertex(160,70){1.3}
\Vertex(70,10){1.3}
\Vertex(100,10){1.3}
\Vertex(130,10){1.3}
\Vertex(160,10){1.3}
\Line[arrow,arrowpos=1,arrowscale=1.5](265,40)(290,40)
\SetOffset(204,10)
\Line(130,70)(160,10)
\Line(100,70)(130,10)
\Line(70,70)(100,10)
\Line[color=white,width=4](160,70)(70,10)
\Line(5,40)(40,40)
\Arc(70,40)(30,90,180)
\Arc(70,40)(30,180,270)
\Line(160,70)(70,10)
\Line(100,70)(70,70)
\Line(130,70)(100,70)
\Line(160,70)(130,70)
\Line(130,10)(160,10)
\Line(100,10)(130,10)
\Line(70,10)(100,10)
\Arc(160,40)(30,270,360)
\Arc(160,40)(30,0,90)
\Line(70,10)(20,0)
\Vertex(40,40){1.3}
\Vertex(190,40){2}
\Vertex(70,70){1.3}
\Vertex(100,70){1.3}
\Vertex(130,70){1.3}
\Vertex(160,70){1.3}
\Vertex(70,10){1.3}
\Vertex(100,10){1.3}
\Vertex(130,10){1.3}
\Vertex(160,10){1.3}
\Text(41,20)[r]{$p$}
\Line[dash,dsize=4](190,40)(220,40)
\Bezier[arrow,arrowpos=1](205,35)(205,5)(180,0)(65,0)
\end{axopicture}
\end{center}
\vspace*{-7mm}
\caption{\small
One external line is moved to create a Feynman diagram that can be
integrated, here done for the topology of fig.~\ref{fig:gluons}. 
One should take into account that there can be up to 5 powers of dot 
products in the numerator, causing many UV subdivergences. 
Furthermore, the double propagator that remains on the right can 
introduce IR divergences. After the subdivergences are subtracted, 
the integral over $p$ can be performed and the remaining four-loop 
topology can be handled by the {\sc Forcer} program.}
\label{fig:rearrange}
\end{figure}

The key property of the $R^*$-operation is that the subtraction terms are of 
lower loop order than the original integral. This is made possible via the 
procedure of \textit{IR-rearrangement}. The IR-rearranged integral is, 
in general, any other propagator integral obtained from the original one by 
re-attaching an external momentum in the diagram. 
This is illustrated in fig.~\ref{fig:rearrange}.
For integrals whose superficial degree of divergence (SDD) is higher than
logarithmic, the SDD is reduced by differentiating it sufficiently
many times with respect to its external momenta, before IR-rearranging it.

The upshot is that the IR-rearranged propagator integrals can be chosen to be 
\textit{carpet integrals}, which correspond to graphs where the external lines 
are connected only by a single edge. 
A carpet integral of $L$ loops can be evaluated as a product of an $(L-1)$ 
loop tensor propagator integral times a known one-loop tensor integral. 
In the case of the five-loop beta function this means that one can effectively 
evaluate the poles of \emph{all} five-loop propagator integrals from the 
knowledge of propagator integrals with no more than four loops.

\paragraph{Definition of the $R^*$-operation.}
More precisely, the $R^*$-operation acting on a Euclidean Feynman graph 
$\Gamma$ can be written as
\beq
R^* (\Gamma) \:= \!\sum_{\substack{\gamma \subseteq \Gamma,\tilde \gamma  
  \subseteq \Gamma \\[0.5mm]\gamma  \cap \tilde \gamma  = \emptyset}}
  \widetilde Z(\tilde \gamma )* Z(\gamma ) * \Gamma/\gamma  \setminus 
  \tilde \gamma 
\:.
\eeq
Here the sum goes over disjoint pairs of UV and IR subgraphs $\gamma $ and 
$\tilde \gamma $ respectively. 
The UV subgraph is defined identically as in the case of the $R$-operation,
a possibly disconnected subgraph whose connected components are superficially 
UV divergent 1PI subgraphs. 
To define the IR subgraph $\tilde \gamma$ is analogous but more involved than 
for UV subgraphs, and is for this reason referred to the literature 
\cite{RSTAR1982,RSTAR1984,RSTAR1985,RSTAR1991,Herzog:2017bjx}. 
The remaining contracted graph $\Gamma/\gamma \setminus\!\tilde \gamma $ 
is constructed by first contracting the components $\gamma$ in $\Gamma$ 
and then deleting the lines and vertices contained in $\tilde \gamma $ in 
$\Gamma/\gamma $. 
The case in which $\tilde \gamma =\Gamma$ can occur only if $\Gamma$ is a 
scaleless vacuum graph of logarithmic superficial degree of divergence. 
In this case $\Gamma\setminus\!\tilde \gamma $ is defined as the unit~$1$. 
The UV and IR counterterm operations $Z$ and $\tilde Z$ are then defined
recursively via
\beq
 Z (\Gamma) \:=\, -K\Big(\sum_{\substack{\gamma  
 \subsetneq \Gamma,\tilde \gamma  \subseteq \Gamma \\[0.5mm]
 \gamma  \cap \tilde \gamma  = \emptyset}}
 \widetilde Z(\tilde \gamma )* Z(\gamma ) * \Gamma/\gamma  
 \setminus \tilde \gamma \Big)
\,,
\eeq
where one omits in the sum over UV subgraphs the full graph $\Gamma$, and
\beq
 \tilde Z (\Gamma_0) \:=\, -K\Big(\sum_{\substack{\gamma 
 \subseteq \Gamma_0,\tilde \gamma  \subsetneq \Gamma_0 \\[0.5mm]
 \gamma  \cap \tilde \gamma  = \emptyset}}
 \widetilde Z(\tilde \gamma )* Z(\gamma ) * \Gamma_0/\gamma  
 \setminus \tilde \gamma \Big)
\,,
\eeq
where one omits in the sum over IR subgraphs the scaleless vacuum Feynman 
graph $\Gamma_0$. The identity $R^*(\Gamma_0)=0$ can be used to find
relations among IR and UV counterterms in dimensional regularisation.

\vspace{1mm}
It is useful to write
\beq
R^*\,=\,\mathrm{id}+\delta R^*\,,
\eeq
with $\delta R^*$ collecting all counterterms and $\mathrm{id}$ the identity
map. From the finiteness of $R^*(\Gamma)$ we then obtain
\beq
K\circ R^*(\Gamma) \,=\,0
\,,
\eeq
i.e., the image of $R^*$ is in the kernel of the pole operator $K$. 
It follows that the pole part of $\Gamma$ is given by
\beq
  K(\Gamma)=-K\circ \delta R^*(\Gamma)\,.
\eeq
In principle, the $Z$ operation would be all one needs to extract the local 
UV divergence, and hence the $\beta$ function. However, the operation $Z$ 
does not commute with algebraic operations on the integrand such as 
contracting the Feynman rules with projectors or taking traces over Dirac 
matrices. 
On the other hand, the dimensionally regulated integrals are of course 
entirely unaffected by such operations. 
This is the main advantage of using $\delta R^*$ over $Z$, see ref.~%
\cite{Herzog:2017bjx} for more details.

\paragraph{Example.}
Let us consider the Feynman integral
\beq
  \Gamma\:=\:
  \bubtwodotnum{1}{2}{3} \:=\: 
  \int 
  \frac{d^Dk_1\,d^Dk_2}{(k_1^2)^2(k_2+P)^2(k_1+k_2)^2}
\eeq
Here we have labeled the lines, such that their corresponding momenta are 
parameterised as $q_1=k_1,q_2=k_2+P,q_3=k_1+k_2$ respectively. 
The example features an IR divergence when the momentum is flowing through 
the dotted line $1$ vanishes. 
It also features two UV divergent subgraphs, corresponding to the full graph 
and the subgraph which consists of lines $2$ and $3$. 
The action of the $R^*$ operation yields
\bea
 R^*\l(\bubtwodotnum{1}{2}{3}\r)&\,=\,&\bubtwodotnum{1}{2}{3}
 +Z\l( \bubtwodotnum{1}{2}{3}\r)
 +Z\l( \bubbleonenum{2}{3}\r)* \tadonenum{1}\\
 &&+\tilde Z\l(\IRonenum{1}\r)*\bubbleonenum{2}{3}
 +\tilde Z\l(\IRonenum{1}\r)*Z\l( \bubbleonenum{2}{3}\r)*1
\,.\nn
\eea
The IR counterterm can be evaluated as
\beq
  \tilde Z\l(\IRonenum{}\r) \,=\,
  \tilde Z\l(\vaconenum{}{}\r) \,=\, 
  -Z\l(\vaconenum{}{}\r)=K\l(\bubbleonenum{}{} \r)
\,.
\eeq
The poles of the integral are then given by
\bea
  K\l(\bubtwodotnum{1}{2}{3}\r) &\,=\,& 
 -K\circ \delta R^*\l(\bubtwodotnum{1}{2}{3}\r) \nn \\ &\,=\,&
 -K\Bigg[
 Z\l( \bubtwodotnum{1}{2}{3}\r)
 +Z\l( \bubbleonenum{2}{3}\r)* \tadonenum{1}\\
 &&+K\l(\bubbleonenum{}{} \r)*\bubbleonenum{2}{3}
 +K\l(\bubbleonenum{}{} \r)*Z\l( \bubbleonenum{2}{3}\r)*1\Bigg]
\,.\nn
\eea
Each of the counterterms on the right hand side are one-loop or carpet-type 
two-loop integrals which can be evaluated straightforwardly.

\subsection{Diagram computations and analysis}

The Feynman diagrams for the background-field propagator up to five loops 
were generated with {\sc Qgraf} \cite{QGRAF}. 
They were heavily manipulated by a {\sc Form} 
\cite{Vermaseren:2000nd,Kuipers:2012rf,Ruijl:2017dtg} program that 
determined the topology and computed the colour factor according to
ref.~\cite{Colour}.
Additionally, it merged diagrams of the same topology, colour factor, and 
maximal power of $\nf$ into meta diagrams for computational efficiency. 
Vanishing integrals containing massless tadpoles or symmetric colour tensors 
with an odd number of indices were filtered out from the beginning. 
Lower-order self-energy insertions were treated as described in 
ref.~\cite{jvLL2016}.
In this manner we arrived at 2 one-loop, 9 two-loop, 55 three-loop, 
572 four-loop and 9414 five-loop meta diagrams.

\vspace{1mm}
The diagrams up to four loops had been computed earlier to all powers of the 
gauge parameter $\xi$ using the {\sc Forcer} program \cite{Ruijl:2017cxj}.  
The five-loop part of our computation was restricted to the Feynman gauge,
$\xi_F^{} = 1 - \xi = 0$. 
An extension to the first power in $\xi_F^{}$ would have been considerably 
slower;
the five-loop computation for a general $\xi$ would have been impossible 
without substantial further optimizations of our code. 
Instead, we verified our computations by checking the relation
$\, Q_\mu Q_\nu\, \Pi_B^{\:\!\mu\nu} \,=\, 0\, $ required 
by eq.~(\ref{eq:PIBtensor}).
This check took considerably more time than the actual determination of the
five-loop beta function.
The later computations in refs.~\cite{Luthe:2017ttg,Chetyrkin:2017bjc},
performed via massive tadpole integrals and the global $R^*$ method, included 
terms linear in $\xi_F^{}$ and all powers of $\xi_F^{}$, respectively. 

\vspace{1mm}
The five-loop diagrams were calculated on computers with a combined
total of more than 500 cores, 80\% of which are older and slower by a 
factor of almost three than the latest workstations we had in 2016. 
One core of the latter performed
a `raw-speed' {\sc Form} benchmark, a four-dimensional trace of 14 Dirac
matrices, in about 0.02 seconds which corresponds to 50 `form units' (fu)
per hour. 
The total CPU time for the five-loop diagrams was $3.8 \cdot 10^{7}$ seconds 
which corresponds to about $2.6 \cdot 10^{5}$ fu on the computers used. 
The {\sc TForm} parallelization efficiency for single meta diagrams run with 
8 or 16 cores was roughly 0.5; the whole calculation of the beta function 
(without the check mentioned above), distributed `by hand' over the available 
machines, finished in three days.
For comparison, a corresponding $R^*$ computation for $\xi_F^{} = 0$ at
four loops required about $10^{3}$ fu, which is roughly the same as for the
first computation of the four-loop beta function to order $\xi_F^{\,1}$ by a
totally different method in ref.~\cite{vanRitbergen:1997va}. 
The computation with the {\sc Forcer} program at four and fewer loops is 
much faster. 

\vspace{1mm}
The determination of $Z_B$ from the unrenormalized background propagator is 
performed by imposing, order by order, the finiteness of its renormalized 
counterpart. 
The beta function can be read off from the $1/\ep$ coefficients of $Z_B$; 
the higher poles of $Z_B$ are fixed by lower-order information and thus 
provide valuable checks. 
If the calculation is performed in the Landau gauge, the gauge parameter does 
not have to be renormalized. In a $k$-th order expansion about the Feynman 
gauge at five loops, the $L\!<\!5$ loop contributions are needed up to
$\xi_F^{\,5-L+k}$. The four-loop renormalization constant for the gauge
parameter is not determined in the background field and has to be `imported'.
In the present $k=0$ case, the terms already specified in 
ref.~\cite{Czakon:2004bu} would have been sufficient had we not performed the
four-loop calculations to all powers of $\xi_F^{}$ anyway.

\newpage

\section{Results for the beta functions}

Here we present the analytic expressions for the beta function of 
gauge theories with a single compact Lie group. These include, 
in particular, the case of QCD with $\nf$ flavours of quarks.
We then address the stability of its perturbative expansion and of 
the resulting running of the renormalized coupling constant in the 
standard \MSb\ scheme. 
See refs.~\cite{Ruijl:2017eht,Ruijl:2018poj}
for the corresponding results in a relatively common alternative,
the minimal momentum subtraction (MiniMOM) scheme 
\cite{vonSmekal:2009ae}.

\subsection{Analytical expressions}
\allowdisplaybreaks[2]

The perturbative expansion of the beta function can be defined as
\beq
\label{as-run}
  \frac{d}{d \ln\mu^2} \left(\frac{\alpha_s}{4\pi}\right)
  \:=\: \beta(\als)
  \:=\: - \sum_{n=0}^\infty \beta_n \left(\frac{\als}{4\pi}\right)^{n+2} ,
\eeq
where $\als$ is the renormalized coupling and $\mu$ is the 
renormalization scale. The coefficients $\beta_n$ up to four loops,
N$^{n=3}$LO, have been known for a long time
\cite{Gross:1973id,Politzer:1973fx,Caswell:1974gg,Jones:1974mm,%
Tarasov:1980au,Larin:1993tp,vanRitbergen:1997va,Czakon:2004bu},
\bea
\label{beta0}
  \beta_{0} & = &  \frac{11}{3}\: \ca \,-\, \frac{4}{3}\: \tf \nf 
\:\: , \\[2mm]
\label{beta1}
  \beta_{1} & = & \frac{34}{3}\: \cas \,-\, \frac{20}{3}\: \ca\, \tf \nf
    - 4\, \cf\, \tf \nf 
\:\: , \\[2mm]
\label{beta2}
  \beta_{2} & = &  \frac{2857}{54}\: \cat 
    \,-\, \frac{1415}{27}\: \cas\, \tf \nf
    \,-\, \frac{205}{9}\: \cf\, \ca\, \tf \nf \:+\: 2\, \cfs\, \tf \nf 
\nn \\[1mm] & & \mbox{\hspn}
    \,+\, \frac{44}{9}\: \cf\, \tfs \nfs 
    \,+\, \frac{158}{27}\: \ca\, \tfs \nfs  
\:\: , \\[3mm]
\label{beta3}
  \beta_{3} & = &
    \caf \left( \, \frac{150653}{486} - \frac{44}{9}\, \z3 \right)   
    \,+\, \dfAAna \left(  - \frac{80}{9} + \frac{704}{3}\,\z3 \right)
\nn \\[0mm] & & \mbox{\hspn}
    \,+\,  \cat\, \tf  \nf
      \left(  - \frac{39143}{81} + \frac{136}{3}\, \z3 \right)
    \,+\, \cas\, \cf\, \tf  \nf
      \left( \, \frac{7073}{243} - \frac{656}{9}\,\z3 \right)
\nn \\[0mm] & & \mbox{\hspn}
    \,+\, \ca\, \cfs\, \tf  \nf
      \left(  - \frac{4204}{27} + \frac{352}{9}\,\z3 \right)
    \,+\, \dfFAna\, \nf \left( \, \frac{512}{9} - \frac{1664}{3}\,\z3 \right)
\nn \\[0mm] & & \mbox{\hspn}
    \,+\, 46\, \cft\, \tf  \nf
    \,+\,  \cas \tfs \nfs 
      \left( \, \frac{7930}{81} + \frac{224}{9}\,\z3 \right)
    \,+\,  \cfs\, \tfs\, \nfs
      \left( \, \frac{1352}{27} - \frac{704}{9}\,\z3 \right)
\nn \\[0mm] & & \mbox{\hspn}
    \,+\,  \ca\, \cf\, \tfs\, \nfs
      \left( \, \frac{17152}{243} + \frac{448}{9}\,\z3 \right)
    \,+\, \dfFFna\, \nfs\, \left( - \frac{704}{9} + \frac{512}{3}\,\z3 \right)
\nn \\[0mm] & & \mbox{\hspn}
    \,+\, \frac{424}{243}\: \ca\, \tft\, \nft 
    \,+\, \frac{1232}{243}\: \cf\, \tft\, \nft  
\:\: .
\end{eqnarray} 
The corresponding five-loop contribution was computed a little less than 
a decade ago, 
first for QCD \cite{Baikov:2016tgj}, 
then for a general gauge group in our main paper \cite{Herzog:2017ohr}, 
and shortly thereafter also in refs.~\cite{Luthe:2017ttg,Chetyrkin:2017bjc}.
The resulting coefficient in eq.~(\ref{as-run}) reads
  
\pagebreak
\vspace*{-1.5cm}
\bea
\label{beta4}
  \beta_{4}^{} & = & 
       \cai \left( 
           {8296235 \over 3888} 
         - {1630 \over 81} \*\, \z3 
         + {121 \over 6} \*\, \z4 
         - {1045 \over 9} \*\, \z5 \right) 
\nn \\[0mm] & & \mbox{\hspn}
       \:+\: \dfAAna\, \* \ca \* \left( 
         - {514 \over 3}  
         + {18716 \over 3} \,\* \z3 
         - 968 \*\, \z4 
         - {15400 \over 3} \*\, \z5 
           \right) 
\nn \\[0mm] & & \mbox{\hspn}
       \:+\: \caf\* \,\tf\* \nf\* \left(  
         - {5048959 \over 972}  
         + {10505 \over 81} \* \,\z3 
         - {583 \over 3} \* \,\z4 
         + 1230\* \,\z5 \right)
\nn \\[0mm] & & \mbox{\hspn}
       \:+\: \cat\* \,\cf\* \,\tf\* \nf\* \left( \,
           {8141995 \over 1944}  
         + 146\* \,\z3 
         + {902 \over 3} \* \,\z4 
         - {8720 \over 3} \* \,\z5 \right)
\nn \\[0mm] & & \mbox{\hspn}
       \:+\: \cas\* \,\cfs\* \,\tf\* \nf\* \left( 
         - {548732 \over 81}  
         - {50581 \over 27} \* \,\z3 
         - {484 \over 3} \* \,\z4 
         + {12820 \over 3} \* \,\z5 \right)
\nn \\[0mm] & & \mbox{\hspn}
       \:+\: \ca\* \,\cft\* \,\tf\* \nf\* \left( 
           3717 
         + {5696 \over 3} \* \,\z3 
         - {7480 \over 3} \* \,\z5 \right)
       \:-\: \cff\* \,\tf\* \nf\* \left( \,
           {4157 \over 6}  
         + 128\* \,\z3 \right)
\nn \\[0mm] & & \mbox{\hspn}
       \:+\: \dfAAna \* \,\tf\* \nf\* \left( 
           {904 \over 9}  
         - {20752 \over 9} \* \,\z3 
         + 352\* \,\z4 
         + {4000 \over 9} \* \,\z5 \right)
\nn \\[0mm] & & \mbox{\hspn}
       \:+\: \dfFAna \* \,\ca\* \,\nf\* \left( \,
           {11312 \over 9}  
         - {127736 \over 9} \* \,\z3 
         + 2288\* \,\z4 
         + {67520 \over 9} \* \,\z5 \right)
\nn \\[0mm] & & \mbox{\hspn}
       \:+\: \dfFAna \* \,\cf\* \,\nf\* \left( 
         - 320 
         + {1280 \over 3} \* \,\z3 
         + {6400 \over 3} \* \,\z5 \right)
\nn \\[0mm] & & \mbox{\hspn}
       \:+\: \cat\* \,\tfs\* \nfs \* \left( \,
           {843067 \over 486}  
         + {18446 \over 27} \* \,\z3 
         - {104 \over 3} \* \,\z4 
         - {2200 \over 3} \* \,\z5 \right)
\nn \\[0mm] & & \mbox{\hspn}
       \:+\: \cas\* \,\cf\* \,\tfs\* \nfs \* \left( \,
           {5701 \over 162}  
         + {26452 \over 27} \* \,\z3 
         - {944 \over 3} \* \,\z4 
         + {1600 \over 3} \* \,\z5 \right)
\nn \\[0mm] & & \mbox{\hspn}
       \:+\: \cfs\* \,\ca\* \,\tfs\* \nfs \* \left( \,
           {31583 \over 18}  
         - {28628 \over 27} \* \,\z3 
         + {1144 \over 3} \* \,\z4 
         - {4400 \over 3} \* \,\z5 \right)
\nn \\[0mm] & & \mbox{\hspn}
       \:+\: \cft\* \,\tfs\* \nfs \* \left( 
         - {5018 \over 9}  
         - {2144 \over 3} \* \,\z3 
         + {4640 \over 3} \* \,\z5 \right)
\nn \\[0mm] & & \mbox{\hspn}
       \:+\: \dfFAna \* \,\tf\* \nfs \* \left( 
         - {3680 \over 9}  
         + {40160 \over 9} \* \,\z3 
         - 832\* \,\z4 
         - {1280 \over 9} \* \,\z5 \right)
\nn \\[0mm] & & \mbox{\hspn}
       \:+\: \dfFFna \* \,\ca\* \,\nfs \* \left( 
         - {7184 \over 3}  
         + {40336 \over 9} \* \,\z3 
         - 704\* \,\z4 
         + {2240 \over 9} \* \,\z5 \right)
\nn \\[0mm] & & \mbox{\hspn}
       \:+\: \dfFFna \* \,\cf\* \,\nfs \* \left( \,
           {4160 \over 3}  
         + {5120 \over 3} \* \,\z3 
         - {12800 \over 3} \* \,\z5 \right)
\nn \\[0mm] & & \mbox{\hspn}
       \:+\: \cas\* \,\tft\* \nft \* \left( 
         - {2077 \over 27}  
         - {9736 \over 81} \* \,\z3 
         + {112 \over 3} \* \,\z4 
         + {320 \over 9} \* \,\z5 \right)
\nn \\[0mm] & & \mbox{\hspn}
       \:+\: \ca\* \,\cf\* \,\tft\* \nft \* \left(  
         - {736 \over 81}  
         - {5680 \over 27} \* \,\z3 
         + {224 \over 3} \* \,\z4 \right)
       \:+\: \ca\* \,\tff\* \nff \* \left( \,
           {916 \over 243}  
         - {640 \over 81} \* \,\z3 \right)
\nn \\[0mm] & & \mbox{\hspn}
       \:+\: \cfs\* \,\tft\* \nft \* \left( 
         - {9922 \over 81}  
         + {7616 \over 27} \* \,\z3 
         - {352 \over 3} \* \,\z4 \right)
       \:-\: \cf\* \,\tff\* \nff \* \left( \,
           {856 \over 243}
         + {128 \over 27} \* \,\z3 \right)
\nn \\[0mm] & & \mbox{\hspn}
       \:+\: \dfFFna \* \,\tf\* \nft\* \left( \,
           {3520 \over 9}  
         - {2624 \over 3} \* \,\z3
         + 256\* \,\z4 
         + {1280 \over 3} \* \,\z5 \right)
\:\: .
\eea
These coefficients are the same in all MS-like schemes, i.e., within the class
of renormalization schemes that differ only by a shift of the scale $\mu$.
For an SU($N$) gauge group and fermions transforming according to its
fundamental representation, the group invariants (`colour factors') in 
eqs.~(\ref{beta0}) -- (\ref{beta4}) are given by
\bea
\label{colSU(N)}
 &&
 \ca \:=\: N
\; , \quad
 \cf \:=\: \frac{N_A}{2 N} \:=\: \frac{N^2-1}{2 N}
\; , \quad
 \dfAAna \:=\: \frac{N^2(N^2+36)}{24}
\; , \nn \\[0mm] && \qquad
 \dfFAna \:=\: \frac{ N(N^2+6)}{48}
\; , \quad
 \dfFFna \:=\: \frac{N^4-6N^2+18}{96\, N^2}
\eea
together with $\tf = 1/2$.
The results for QED (i.e., the group U(1)) are obtained for $\ca=0$, 
$d_A^{\,abcd}=0$, $\cf = 1$, $\tf = 1$, $d_F^{\,abcd}=1$, and $N_A = 1$, 
see refs.~\cite{betaQED1,Baikov:2012zm,Herzog:2017ohr}.
The~reader is referred to ref.~\cite{vanRitbergen:1997va} for a discussion 
of other gauge groups.


\subsection{Numerical consequences}

Inserting the numerical values of the Riemann zeta function,
$\z3 \cong 1.2020569$, $\z4 = \pi^4/90 \cong 1.082323$ and
$\z5 \cong 1.0369278$, the normalized beta function of QCD,
$\widetilde{\beta} \equiv - \beta(\als) / (\ar(2) \beta_0)$ 
with $\als = 4 \pi\, a_{\rm s}$, is found to be
\bea
\label{bqcd-num}
  \;\widetilde{\beta}(\als,\nf\!=\!3) & = &
       1
       + 0.56588 \,\* \als
       + 0.45301 \,\* \as(2)
       + 0.67697 \,\* \as(3)
       + 0.58093 \,\* \as(4)
       + \dots
\, , \quad\; \nn \\[0mm]
  \;\widetilde{\beta}(\als,\nf\!=\!4) & = &
       1
       + 0.49020 \,\* \als
       + 0.30879 \,\* \as(2)
       + 0.48590 \,\* \as(3)
       + 0.28060 \,\* \as(4)
       + \dots
\, , \quad\; \nn \\[0mm]
  \;\widetilde{\beta}(\als,\nf\!=\!5) & = &
       1
       + 0.40135 \,\* \als
       + 0.14943 \,\* \as(2)
       + 0.31722  \,\* \as(3)
       + 0.08092  \,\* \as(4)
       + \dots
\, , \quad\; \nn \\[0mm]
  \;\widetilde{\beta}(\als,\nf\!=\!6) & = &
       1
       + 0.29557  \,\* \als
       - 0.02940  \,\* \as(2)
       + 0.17798  \,\* \as(3)
       + 0.00156  \,\* \as(4)
       + \dots
\eea
for the physically relevant values of $\nf$. 
In contrast to $\beta_0$, $\beta_1$, and $\beta_2$, which change sign at 
about $\nf = 16.5$, 8.05, and 5.84 respectively, $\beta_3$ and $\beta_4$ 
are positive (except at very large $\nf$ for $\beta_4$), but have (local) 
minima at $\nf \simeq 8.20$ and $\nf \simeq 6.07$. 

\begin{figure}[htb]
\vspace{-2mm}
\centerline{\epsfig{file=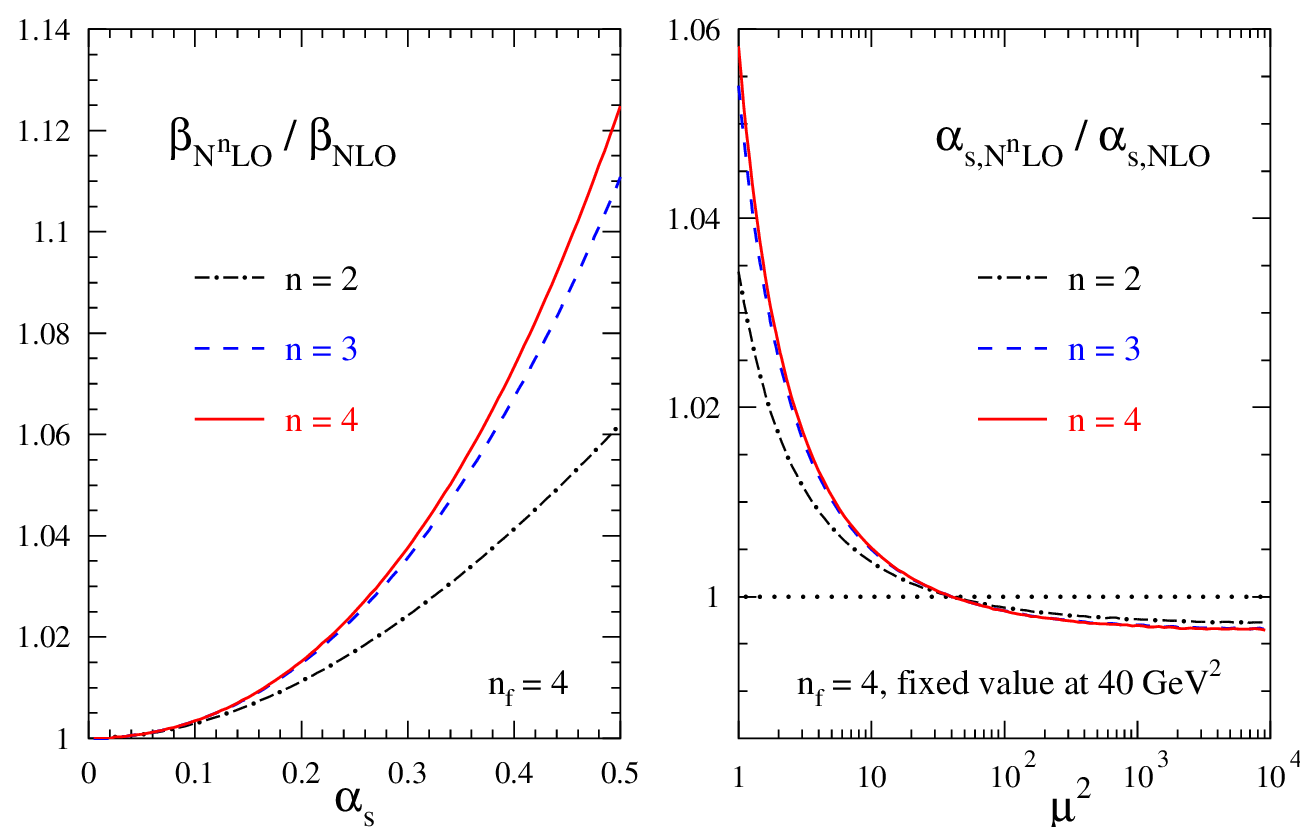,width=12.0cm,angle=0}}
\vspace{-3mm}
\caption{ \label{fig:asrun} \small
 The N$^2$LO, N$^3$LO and N$^4$LO results for the beta function of QCD for 
 four flavours, and the resulting running of $\als$ for a fixed value of 
 0.2 at $40 \mbox{ GeV}^2$.  All curves are normalized to the NLO results 
 in order to show the higher-order effects more clearly.} 
\vspace{-1mm}
\end{figure}

\vspace{1mm}
These results are illustrated in fig.~\ref{fig:asrun} for $\nf = 4$. 
An order-independent value of $\als = 0.2$ at $\mu^2 = 40 \mbox{ GeV}^2$ 
has been chosen in order to only show the
differences caused by the beta-function.
A realistic order dependence of $\als$ at this scale, as determined from the 
scaling violations in inclusive deep-inelastic scattering, would be 0.208, 
0.201, 0.200, and 0.200, respectively, at N$^n$LO for $n=1,\,2,\,3,\,4$
\cite{Vermaseren:2005qc}.

\vspace{1mm}
Including the N$^4$LO term changes $\beta(\als)$ by less than 1\%
at $\als\!\leq\! 0.47$ for $\nf = 4$ and at $\als \leq 0.39$ for 
$\nf = 3$; the corresponding values at N$^3$LO are significantly 
smaller with 0.29 and 0.26. 
The N$^4$LO effect on the values of $\als$ as shown in 
fig.~\ref{fig:asrun} are as small as 0.08\% (0.4\%) at 
$\mu^2 = 3 \mbox{ GeV}^2$ ($1 \mbox{ GeV}^2$); 
the corresponding N$^3$LO corrections are larger by about a factor of 5.

\vspace{1mm}
In order to further illustrate the perturbative behaviour of the beta 
functions of QCD and pure ($\nf = 0$) SU($N$) Yang-Mills theories, 
one can use the quantities
\beq
\label{ashat}
  \widehat{\alpha}_{\rm s}^{\,(n)}(\nf) \; = \; 4 \pi\,
  \left|\, \frac{\beta_{n-1}(\nf)}{4\,\beta_{n}(\nf)} \,\right| 
\:\: , \quad  
  \widehat{\alpha}_{\rm YM}^{\,(n)}(N) \; = \; 4 \pi\, N
  \left|\, \frac{\beta_{n-1}(N)}{4\,\beta_{n}(N)} \,\right| \:\: .
\eeq
Recalling the normalization (\ref{as-run}) of our expansion parameter,
$\widehat{\alpha}_{\rm s}^{\,(n)}(\nf)$ represents the value of $\als$ for
which the $n$-th order correction is 1/4 of that of the previous order.
Hence $\als \lsim \widehat {\alpha}_{\rm s}^{\,(n)}(\nf)$ defines
(somewhat arbitrarily due to the choice of a factor of 1/4) a region of 
fast convergence of $\beta(\als,\nf)$.
As the absolute size of the $n$-th and $(n\!-\!1)$-th order effects are 
equal for $\als= 4\,\widehat{\alpha}^{\,(n)}(\nf)$, the quantity 
(\ref{ashat}) also indicates where the expansion appears not to be 
reliable anymore, $\als \gsim 4\,\widehat{\alpha}_{\rm s}^{\,(n)}(\nf)$,
for values of $\nf$ that are not too close to zeros or minima of 
$\beta_{n-1}$ and $\beta_n$.

\vspace{1mm}
The factor $N$ in $\widehat{\alpha}_{\rm YM}^{\,(n)}(N)$ compensates the 
leading large-$N$ dependence $N^{n+1}$ of $\beta_n$. The parameter 
that needs to be small in SU($N$) Yang-Mills theory is thus not 
$\alpha_{\rm YM}^{}$, but $N \alpha_{\rm YM\,}^{}$.

\begin{figure}[thb]
\vspace{-2mm}
\centerline{\epsfig{file=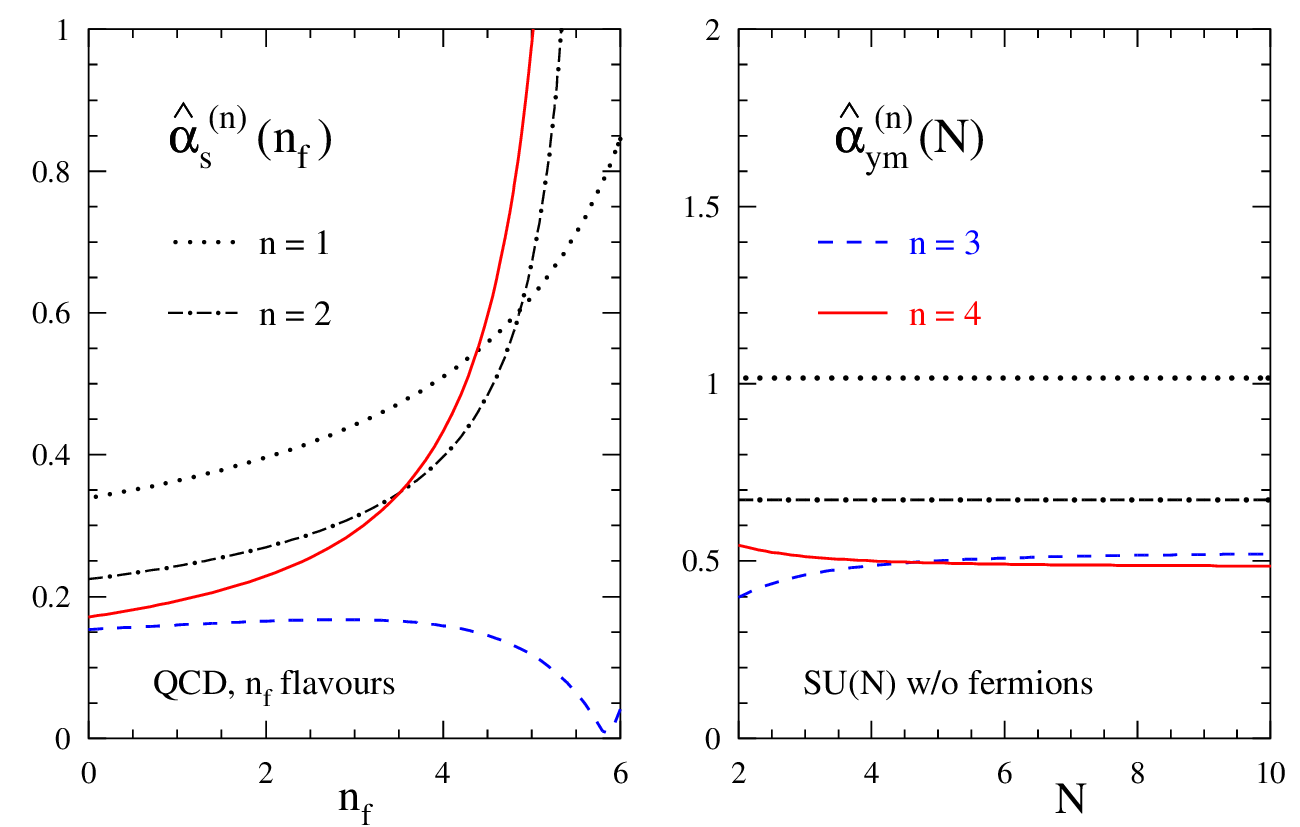,width=12.0cm,angle=0}}
\vspace{-3mm}
\caption{ \label{fig:ashat} \small
 The values (\ref{ashat}) of the coupling constants
 of QCD and pure SU($N$) Yang-Mills theory for which the absolute
 size of the N$^n$LO contribution to $\beta(\als)$ is a quarter of that of
 the N$^{n-1}$LO term for $n = 1$, 2, 3 (dashed curves) and 4 (solid curves).
 }
\vspace{-1mm}
\end{figure}

\vspace{1mm}
The quantities (\ref{ashat}) are displayed in fig.~\ref{fig:ashat}. 
The behaviour of
$\widehat{\alpha}_{\rm s}^{\,(n)}$ at the upper end of the $\nf$ range
shown in the figure is affected by the zeros and minima of the 
coefficients $\beta_{n}$ mentioned below eq.~(\ref{bqcd-num}).
The $N$-dependence of $\widehat{\alpha}_{\rm YM}$ for pure Yang-Mills theory,
where only terms with $N^{n+1}$ and $N^{n-1}$ enter $\beta_{n}$ (the latter
only at $n \geq 4$ via $d_A^{\,abcd}d_A^{\,abcd}/N_A$, 
cf.~eq.~(\ref{colSU(N)}) above), is rather weak. 

\vspace{1mm}
With only the curves up to four loops, one may have been tempted 
to draw conclusions from the substantial shrinking of the `stable' 
$\als$ region from NLO to N$^2$LO and from N$^2$LO to N$^3$LO that 
are not supported by the N$^4$LO (five-loop) results: 
this shrinking does not continue, and is even reversed in QCD for 
the physically relevant values of $\nf$, from N$^3$LO to N$^4$LO.

\newpage

\section{Further five-loop computations}

The methods developed for the determination of the five-loop beta 
function can be extended to other, computationally even more
demanding cases.

\vspace{0.5mm}
At first surprisingly, these cases include certain decay rates of 
the Higgs boson and the hadronic $R$ ratio, 
$R = \sigma_{e^+e^- \ra \: \mbox{\footnotesize hadrons}} 
\: / \sigma_{e^+e^- \ra \: \mu^+\mu^-}$. 
All these involve imaginary parts of self-energies, which 
can be obtained by analytic continuations
\beq
\label{AnCont}
  \mbox{Im } \Pi(-q^2-i\delta) 
  \:=\: \mbox{Im } e^{i \pi \ep L\,} \Pi(q^2)
  \:=\: \sin(L\pi \ep)\, \Pi(q^2)
\;,
\eeq
where $\ep=\frac{1}{2}\,(4-D)$ is the dimensional regulator and 
$L$ the number of loops. The crucial point is now that these imaginary 
parts are suppressed by a factor of $\ep\:\!$:
\beq
  \sin(L\pi \ep) \:=\: L \pi \ep \big(1 - \frct{1}{3!}\, (L \pi \ep)^2
  + \frct{1}{5!}\, (L \pi \ep)^4 + \,\ldots \big)
\; .
\eeq
Therefore the finite parts of $\mbox{Im } \Pi(-q^2)$ can be obtained 
from the $1/\ep$ term of $\Pi(q^2)$ which in turn can be computed via 
the $R^*$-operation. Below we report on Higgs decay to gluons in
the heavy-top limit; for results on $H \ra b \bar{b}$ and the 
hadronic $R$-ratio see ref.~\cite{Herzog:2017dtz} and references 
therein.

\vspace{0.5mm}
A conceptually more straightforward application of the $R^*$
operation is the determination of low-$N$ Mellin moments of the
N$^4$LO splitting functions for the scale dependence of non-singlet
combinations of the quark distributions in hadrons, which can be 
obtained from the $1/\ep$ pole terms of five-loop operator matrix
elements. The results for $N=2$ and $N=3$ were obtained in 
ref.~\cite{Herzog:2018kwj};
already for $N=4$ the hardest Feynman diagrams were too demanding
at that time in terms of run time and required disk space for 
the intermediate expressions.

\subsection{N$^4$LO Higgs decay to gluons}

In the limit of a heavy top quark and $\nl$ effectively massless 
flavours, the decay of the Higgs boson to hadrons (`to gluons', 
the only leading-order contribution) is related by the optical theorem 
to the imaginary part of the Higgs self-energy,
\beq
\label{GamHgg}
  \Gamma_{\!H\to\, gg}^{} \:=\: 
  \frac{\sqrt{2}\, G_{\rm F}}{\MH}\: |C_1|^2 \,
  \mathrm{Im}\, \Pi^{GG}(-\MHs-i\delta)
\; .
\eeq
Here $G_{\rm F}$ denotes the Fermi constant and $\MH$ the Higgs mass.
The Wilson coefficient includes the dependence on the definition and
value of the top-quark mass $M_t$. It~is known to N${}^4$LO at all 
renormalization scale $\mu$ for the scale invariant (SI), \MSb\ and 
on-shell (OS) top-quark masses 
\cite{C1-2005a,C1-2005b,Chetyrkin:2016uhw,Herzog:2017dtz}.

\vspace{1mm}
After the computation of the Feynman diagrams, the extraction of the
absorptive part and its renormalization, the coefficients $g_n$ up to 
N$^4$LO in
\bea
\label{ImGGexp}
  \frac{4\pi}{N_{\!A}\:\!q^4}\, \mbox{Im}\, \Pi^{\,GG}(q^2) \;\equiv\; G(q^2)
  \:=\:  1 + \sum_{n=1} g_n^{} \ar(n)
\; ,
\eea
lead to the following numerical expansion of $G(q^2)$:
\bea
\label{HggNumNf}
  \nl = 1 &\! : \,&
       1 + 5.437794\,\als   + 20.72031\,\as(2)
         + 58.9218 \,\as(3) +  118.008\,\as(4) + \ldots
\; , \nn \\[0.5mm]
  \nl = 3 &\! : \,&
       1 + 4.695071\,\als   + 13.47244\,\as(2)
         + 20.6639 \,\as(3) - 15.9624 \,\as(4) + \ldots
\; , \nn \\[0.5mm]
  \nl = 5 &\! : \,&
       1 + 3.952348\,\als   + 6.955514\,\as(2)
         - 6.85175 \,\as(3) - 75.2591 \,\as(4) + \ldots
\; , \nn \\[0.5mm]
  \nl = 7 &\! : \,&
       1 + 3.209625\,\als   + 1.169536\,\as(2)
         - 24.4579 \,\as(3) - 76.9977 \,\as(4) + \ldots
\; , \nn \\[0.5mm]
  \nl = 9 &\! : \,&
       1 + 2.466902\,\als   - 3.885496\,\as(2)
         - 32.9870 \,\as(3) - 37.3025 \,\as(4) + \ldots
\;\quad
\eea
at the standard choice $\mu^2 = q^2$ of the renormalization scale
for QCD with up to 5 quark families, i.e., $\nl = 1,\,\ldots,9$
light flavours.
The analytic expressions for a general gauge group
and the generalization to $\mu^2 \neq q^2$ can be found in 
ref.~\cite{Herzog:2017dtz}.

\vspace{1mm}
The effect of the fourth-order correction is larger than that of the previous
order for $\als \gsim 0.1$ in the only physically relevant case of $\nl = 5$.
It is clear from eqs.~(\ref{HggNumNf}), though, that this is not a
generic feature of the QCD perturbation series, but a consequence of the
`accidentally' small size, caused by a sign change close by, of the 
third-order term for this number of flavours. 
A similar situation has been observed for Higgs decay to bottom quarks, 
see refs.~\cite{Baikov:2005rw,Herzog:2017dtz}.

\begin{figure}[thb]
\vspace*{-2mm}
\centerline{\epsfig{file=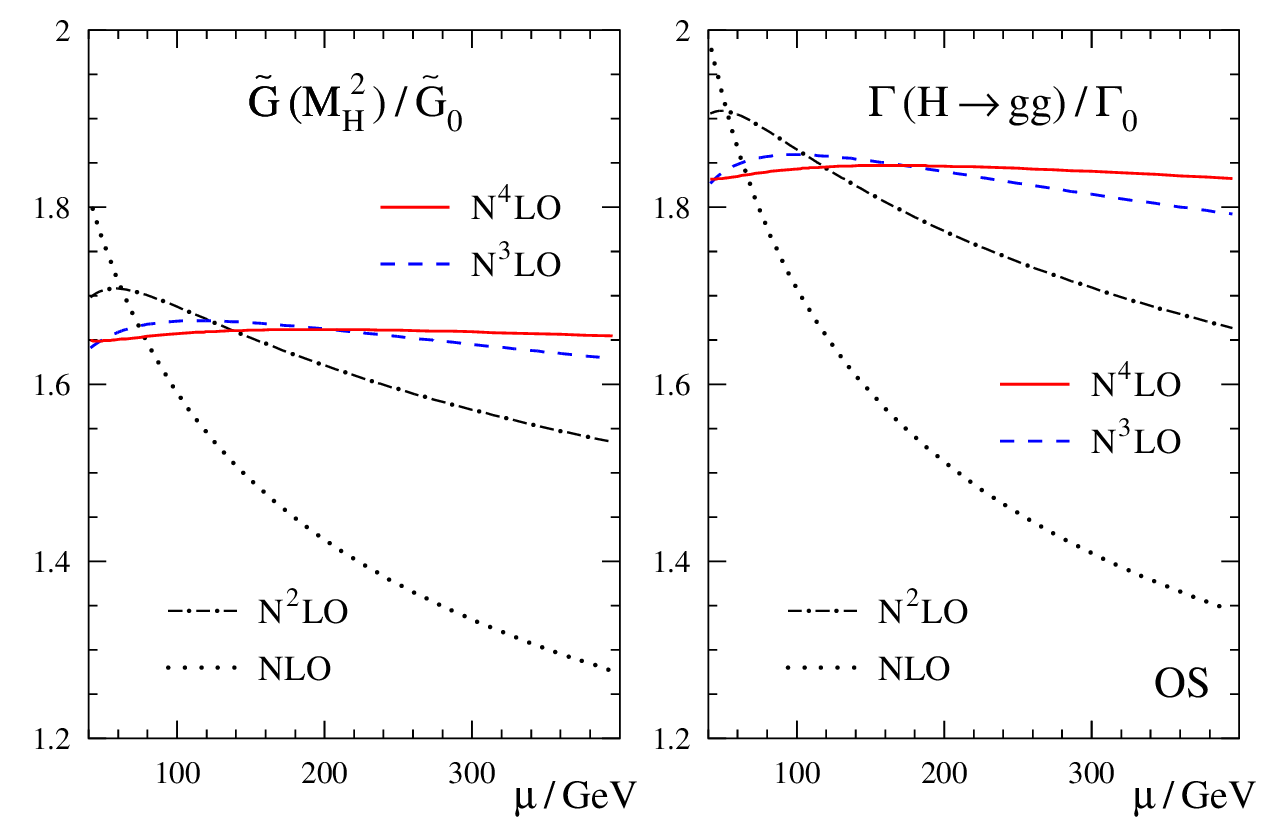,width=12cm,angle=0}}
\vspace{-3mm}
\caption{ \label{fig:hggscale} \small
 The dependence of $\widetilde{G} = (\beta(\ars)/\ars)^2 G(\MHs)$ at $\nl=5$,
 see eq.~(\ref{ImGGexp}) and of the normalized decay width
 $\Gamma_{H \ra\, gg}/\Gamma_0$ on the renomalization scale up to N$^4$LO 
 in the \MSb\ scheme for $\als(\MZs) = 0.118$, $\MH = 125 \mbox{ GeV}$ and 
 an on-shell mass $M_t = 173 \mbox{ GeV}$.}
\vspace{-2mm}
\end{figure}

\vspace{1mm}
The decay rate $\Gamma_{H \ra\, gg}$ in the limit of a heavy top
quark and $\nl$ effectively massless flavours is obtained by combining
eqs.~(\ref{HggNumNf}) with the corresponding expansion of the coefficient 
function $C_1$. 
The resulting $K$-factors, defined by $\Gamma\,=\, K \Gamma_{\mathrm{Born}}$ 
at $\mu^2=\MHs$ read, for an on-shell top mass of $M_t = 173$ GeV,
\bea
\label{KOSnumNf}
  K_{\rm os^{}}(\nl\!=\!1) &\! =\! &
       1 + 7.188498\,\als    + 32.61874\,\as(2)
         + 112.031 \,\as(3)  + 300.278 \,\as(4) + \ldots
\, , \nn \\[0.5mm]
  K_{\rm os^{}}(\nl\!=\!3) &\! =\! &
       1 + 6.445775\,\als    + 23.69992\,\as(2)
         + 56.1329 \,\as(3)  + 64.5259 \,\as(4) + \ldots
\, , \nn \\[0.5mm]
  K_{\rm os^{}}(\nl\!=\!5) &\! =\! &
       1 + 5.703052\,\als    + 15.51204\,\as(2)
         + 12.6660 \,\as(3)  - 69.3287 \,\as(4) + \ldots
\, , \nn \\[0.5mm]
  K_{\rm os^{}}(\nl\!=\!7) &\! =\! &
       1 + 4.960329\,\als    + 8.055116\,\as(2)
         - 19.2021 \,\as(3)  - 120.458 \,\as(4) + \ldots
\, , \nn \\[0.5mm]
  K_{\rm os^{}}(\nl\!=\!9) &\! =\! &
       1 + 4.217606\,\als    + 1.329135\,\as(2)
         - 40.3039 \,\as(3)  - 107.042 \,\as(4) + \ldots
\;\quad
\eea
Corresponding results for a SI top mass of $\mu_t = 164$ GeV 
and the generalization to all renormalization scales can be found in 
ref.~\cite{Herzog:2017dtz}.  
The expansion coefficients in eqs.~(\ref{HggNumNf}) and (\ref{KOSnumNf})
are much larger than those for the beta function above and the 
splitting-function moments below. However, since they are practically 
required only at high scales, and thus for small values of $\als$, 
due to $\MH = 125 \mbox{ GeV}$, the perturbation series are very 
well-behaved as illustrated in fig.~\ref{fig:hggscale}.

\vspace{1mm}
The effect of the N$^4$LO correction to $\Gamma_{H\ra\, gg}$ is $-0.6\%$ at
$\mu = \MH$, and $-0.8\% \:/\, +0.9\%$ at $\mu = 0.5 \:/\, 2\, \MH$,
respectively. The total N$^4$LO result at $\mu = \MH$ is 1.846$\,\Gamma_0$,
and its range in the above scale interval is (1.836 $-$ 1.847)$\,\Gamma_0$.
The N$^4$LO scale variation between $\mu = 1/3\,\MH$ and $\mu = 3\,\MH$
is as small as 0.8\% (full width), a reduction of about a factor of four
with respect to the N$^3$LO result. 
These results are very similar to those for a scale-invariant top mass of 
$\mu_t = 164 \mbox{ GeV}$ \cite{Herzog:2017dtz}.
The dependence of $\Gamma_{\!H\ra\, gg}$ on the top mass is very small,
its largest remaining uncertainty is due to $\als$: 
changing $\als(\MZs)$ by 1\% changes the result by 2.5\%.

\subsection{Low moments of N$^4$LO non-singlet splitting functions}

Via calculations of operator matrix elements,
the odd-N and even-N moments can be determined, respectively, 
for the splitting functions $P_{\rm ns}^{\,+}$ and $P_{\rm ns}^{\,-}$
governing the evolution of flavour differences of quark-antiquarks sums 
($+$) and differences ($-$),
\beq
\label{gpmExp}
  \gamma_{\,\rm ns}^{\:\rm a}(N,\als)
  \:=\: - \int_0^1\! dx \, x^{n-1} P_{\rm ns}^{\:\rm a}(x,\als)
  \:=\: \sum_{n=0} \gamma_{\rm ns}^{(n)\rm a}(N) 
        \left( \frac{\als}{4\pi} \right)^{n-1}
  \, .
\eeq
The first moment of $P_{\rm ns}^{\,-}$ vanishes to all orders,
the respective lowest non-vanishing moments of $P_{\rm ns}^{\,\pm}$
have been computed to five loops in ref.~\cite{Herzog:2018kwj}.

\vspace{1mm}
Combining these results with the lower-order coefficients in 
eq.~(\ref{gpmExp}), see refs.~%
\cite{Velizhanin:2011es,Velizhanin:2014fua,Baikov:2015tea,Moch:2017uml}
and references therein, one arrives at the numerical QCD expansions
\bea
\label{gnspN2}
  \gamma_{\rm ns}^{\,+}(2, \nf\!=\!0) &\!\!=\!\!&
    \gamma_{0}^{}(2) ( 1 + 1.0187\,\as() + 1.5307\, \as(2)
  + 2.3617\, \as(3)  + 4.520 \,\as(4) + \ldots )
\; , \nn \\[-1mm] &\cdots& \nn \\[-1mm]
  \gamma_{\rm ns}^{\,+}(2,\nf\!=\!3) &\!\!=\!\!&
    \gamma_{0}(2) ( 1 + 0.8695\,\as() + 0.7980\, \as(2)
  + 0.9258\,\as(3)  + 1.781 \,\as(4) + \ldots )
\; , \nn \\
  \gamma_{\,\rm ns}^{\,+}(2,\nf\!=\!4) &\!\!=\!\!&
    \gamma_{0}^{}(2) ( 1 + 0.7987\, \as() + 0.5451\, \as(2)
  + 0.5215\: \as(3)  + 1.223 \,\as(4) + \ldots )
\; , \nn \\
  \gamma_{\rm ns}^{\,+}(2,\nf\!=\!5) &\!\!=\!\!&
    \gamma_{0}^{}(2) ( 1 + 0.7280\,\as() + 0.2877\, \as(2)
  + 0.1571\: \as(3)  + 0.849 \,\as(4) + \ldots )
\;\; \nn \\ & & 
\eea
 
\vspace{-2mm}
\noindent
with $\gamma_{0}^{}(2) = 0.28294\,\as()$ at $N\!=\!2$ and
\bea
\label{gnsmN3}
  \gamma_{\,\rm ns}^{\,-}(3,\nf\!=\!0) &\!\!=\!\!&
    \gamma_{0}^{}(3) ( 1 + 1.0153\,\as() + 1.4190\, \as(2)
  + 2.0954\, \as(3)  + 3.954 \,\as(4) + \ldots )
\; , \nn \\[-1mm] &\cdots& \nn \\[-1mm]
  \gamma_{\,\rm ns}^{\,-}(3,\nf\!=\!3) &\!\!=\!\!&
    \gamma_{0}^{}(3) ( 1 + 0.7952\,\as() + 0.7183\, \as(2)
  + 0.7607\, \as(3)  + 1.508 \,\as(4) + \ldots )
\; , \nn \\
  \gamma_{\,\rm ns}^{\,-}(3,\nf\!=\!4) &\!\!=\!\!&
    \gamma_{0}^{}(3) ( 1 + 0.7218\,\as() + 0.4767\, \as(2)
  + 0.3921\, \as(3)  + 1.031 \,\as(4) + \ldots )
\; , \nn \\
  \gamma_{\,\rm ns}^{\,-}(3,\nf\!=\!5) &\!\!=\!\!&
    \gamma_{0}^{}(3) ( 1 + 0.6484\,\as() + 0.2310\, \as(2)
  + 0.0645\, \as(3)  + 0.727 \,\as(4) + \ldots )
\;\; \nn \\ & & 
\eea
  
\vspace{-2mm}
\noindent
with $\gamma_{0}^{}(3) = 0.44210$ at $N\!=\!3$ in the \MSb\ scheme for 
the default choice $\muf\! =\! \mu$ of the factorization scale.
Eqs.~(\ref{gnspN2}) and (\ref{gnsmN3}) include $\nf\! =\! 0$ besides 
the physically relevant values, as it provides further information 
about the behaviour of the series. 
The new N$^4$LO coefficients are larger than one may have expected
from the previous orders. 
This is mostly due to the $\nf$ independent $d_A^{\,abcd}d_A^{\,abcd}$ 
contribution which is large and only enters from this order, see the 
discussion in~ref.~\cite{Herzog:2018kwj}.

\begin{figure}[thb]
\vspace{-1mm}
\centerline{\epsfig{file=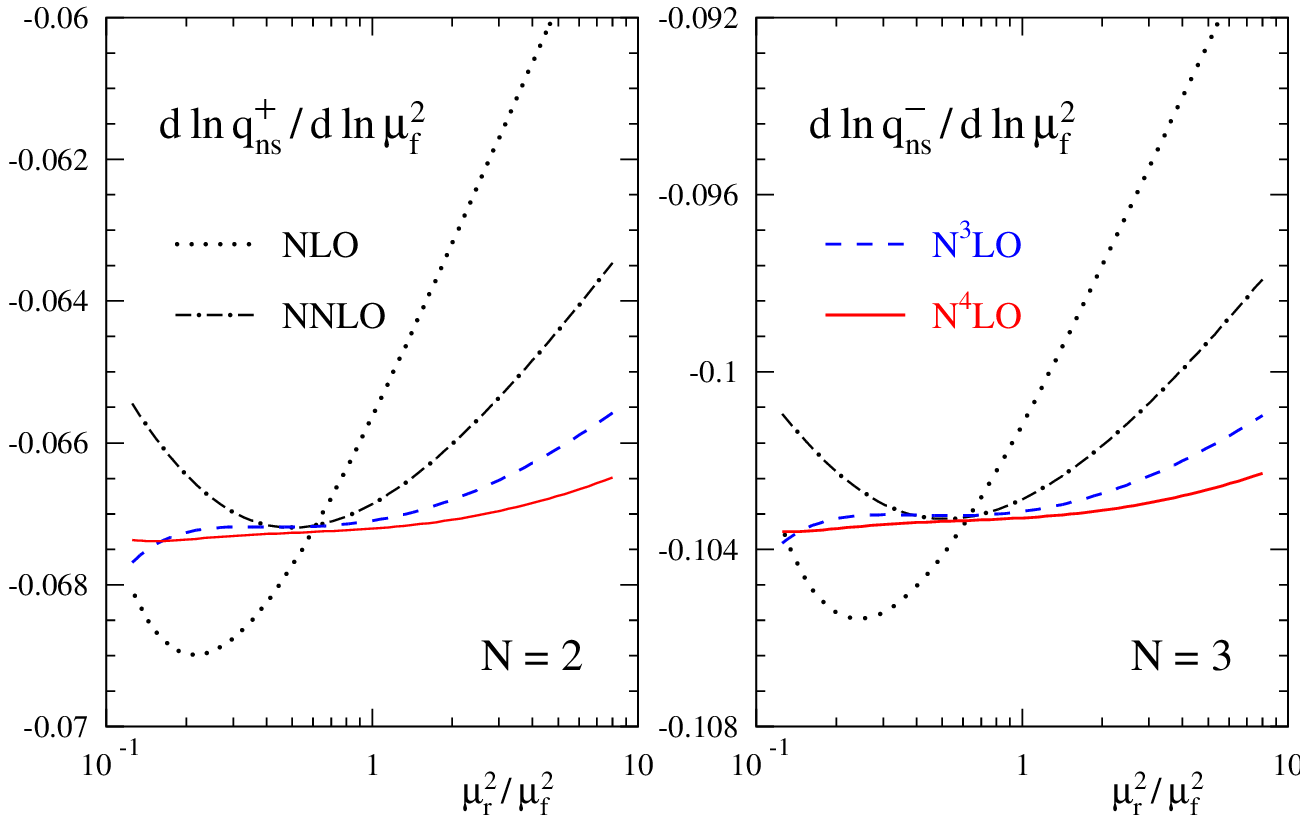,width=12.0cm,angle=0}}
\vspace{-4mm}
\caption{ \label{fig:qnsN23} \small
The renormalization-scale dependence of the logarithmic
factorization-scale derivatives of the quark distributions
$q_{\rm ns}^{\,+}$ at $N=2$ and $q_{\rm ns}^{\,-}$ at $N=3$ at a
standard reference point with $\als(\mufs) = 0.2$ and $\nf=4$.
}
\vspace*{-2mm}
\end{figure}

\vspace{1mm}
The numerical impact of the higher-order contributions to the
anomalous dimensions $\gamma_{\rm ns}^{\,\pm}$ on the evolution of
the $N=2$ and $N=3$ moments of the respective quark distributions
are illustrated in fig.~\ref{fig:qnsN23}.
At $\als(\mufs) = 0.2$ and $\nf=4$, the N$^4$LO corrections
are about 0.15\% for $\mu = \muf$, roughly half the size 
of their N$^3$LO counterparts.
Varying $\mu$ up and down by a factor of 2 leads to a band with a 
full width of about 0.7\%.
The N$^3$LO and N$^4$LO corrections are about twice as large
at a lower scale with $\nf=3$ and $\als(\mufs) = 0.25$.

\vspace{1mm}
The above results have an important application beyond the evolution of quark 
distributions: the leading large-$N$ coefficient of $\gamma_{\rm ns}^{\pm}(N)$ 
identical to the (light-like) quark cusp anomalous dimension $A_{\rm q}$,
a quantity that occurs in numerous other contexts. 
It is known to four loops, see refs.~\cite{Henn:2019swt,vonManteuffel:2020vjv} 
and references therein. 
Using the above results and other information it is possible to obtain a rough 
estimate of the five-loop contribution that leads to
($A_{\rm q,0}^{} = 0.42441\:\as()$)
\bea
\label{AqExpQ}
  A_{\rm q}^{}(\nf\!=\!3) / A_{\rm q,0}^{} &=&
    1  +  0.7266\,\as() +  0.7341\,\as(2) + 0.6647\,\as(3)
       +  ( 1.3 \pm 0.4) \as(4)  +  \ldots 
\: , \nn \\
  A_{\rm q}^{}(\nf\!=\!4) / A_{\rm q,0}^{} &=&
    1  +  0.6382\,\as()  +  0.5100\,\as(2) + 0.3168\,\as(3)
       +  ( 0.8 \pm 0.4 ) \as(4) +  \ldots  
\: , \nn \\
  A_{\rm q}^{}(\nf\!=\!5) / A_{\rm q,0}^{} &=&
    1  +  0.5497\,\as()  +  0.2840\, \as(2) + 0.0133\,\as(3)
       +  ( 0.5 \pm 0.4 ) \as(4)  + \ldots  
\: . \nn \\[-1mm] & &
\eea
 
\vspace{-2mm}
\noindent
See ref.~\cite{Herzog:2018kwj} for more details and a more precise
result in the large-$\nc$ limit.

\newpage
\section{Summary}

We have provided an overview over our computation and result for
the five-loop (N$^4$LO) contributions to the beta function for 
Yang-Mills theories with fermions and Quantum Electrodynamics
\cite{Herzog:2017ohr}. 
A large amount of work went into developing and debugging
a new diagram-by-diagram implementation of the R$^*$ operation,
before we were able --- in just three days, thanks to the considerable
computing resources we had at our disposal then --- to perform
the required diagram computations in the background-field method.
Optimizing our codes further, we were able to carry out further
five-loop calculations \cite{Herzog:2017dtz,Herzog:2018kwj} that were 
computationally much more demanding, mostly due to the higher tensor 
ranks of the Feynman integrals. We have also briefly discussed
the main results of these articles above.  

\vspace{1mm}
Considering the numerical N$^4$LO QCD results in eqs.~(\ref{bqcd-num})
and  (\ref{HggNumNf}) -- (\ref{gnsmN3}), 
together with their implications in figs.~3--6, we conclude that the
expansion in powers of the coupling constant to N$^4$LO is reliable and
provides highly accurate results. While the coefficients relevant to Higgs
decay are much larger than those for the beta function and for the 
low splitting-function moments, they are larger at all orders
in a manner that the results still improve order by order, leading
to a perfectly adequate accuracy at the high scales relevant to actual
physics analyses.

\vspace{1mm}
While having exact (i.e., not only numerical) and general (i.e., not 
only QCD) results as in eqs.~(\ref{beta0}) -- (\ref{beta4}) 
is not relevant to collider physics analyses, it facilitates gaining 
new insights into the mathematical structure of the theory, see, e.g., 
refs.~\cite{Baikov:2018wgs,Baikov:2019zmy,Gracey:2023hzm},
and into formal properties of also other Yang-Mills theories such as 
possible conformal, infrared or ultraviolet fixed points, see refs.~%
\cite{Ryttov:2017kmx,Ryttov:2017dhd,DeCesare:2021pfb}.

\newpage

\subsection*{Acknowledgements}
 
The research reported in ref.~\cite{Herzog:2017ohr} profited from discussions 
with K. Chetyrkin and E. Panzer. 
It was supported by the {\it European Research Council}$\,$ (ERC)
Advanced Grant 320651, {\it HEPGAME} and the UK {\it Science \& Technology
Facilities Council}$\,$ (STFC) grant ST/L000431/1.
We had the opportunity to use most of the {\tt ulgqcd} computer cluster in 
Liverpool which was funded by a far earlier STFC grant, ST/H008837/1, and 
was kept running by S. Downing.
We thank the International Congress of Basic Science (ICBS) for having 
selected our article~\cite{Herzog:2017ohr} for a 2025 Frontiers of Science 
Award in theoretical physics.

\vspace*{-2mm}
\providecommand{\href}[2]{#2}\begingroup\raggedright\endgroup


\address{Higgs Centre for Theoretical Physics, 
  \\School of Physics and Astronomy, The University of Edinburgh\\
  Edinburgh EH9 3FD, Scotland, UK\\
  \email{fherzog@ed.ac.uk}}

\address{Ruijl Research\\
  Chamerstrasse 117, 6300 Zug, Switzerland\\
  \email{benruyl@gmail.com}}

\address{Faculty of Medicine, Juntendo University\\
  1-1 Hiraga-gakuendai, Inzai, Chiba 270-1695, Japan\\
  \email{t.ueda.od@juntendo.ac.jp}}

\address{Nikhef Theory Group\\
  Science Park 105, 1098 XG Amsterdam, The Netherlands\\
  \email{t68@nikhef.nl}}

\address{Department of Mathematical Sciences, University of Liverpool\\
  Liverpool L69 3BX, UK\\
  \email{Andreas.Vogt@liverpool.ac.uk}}

\end{document}